# The Origin and Non-quasiparticle Nature of Fermi Arcs in $Bi_2Sr_2CaCu_2O_{8+\delta}$

T. J. Reber[1], N. C. Plumb[1], Z. Sun[1], Y. Cao[1], Q. Wang[1], K. McElroy[1], H. Iwasawa[2], M. Arita[2], J. S. Wen[3], Z. J. Xu[3], G. Gu[3], Y. Yoshida[4], H. Eisaki[4], Y. Aiura[4], and D. S. Dessau[1]

[1]*Dept. of Physics, University of Colorado, Boulder, 80309-0390, USA*

[2]*Hiroshima Synchrotron Radiation Center, Hiroshima University, Higashi-Hiroshima, Hiroshima 739-0046, Japan*

[3] *Condensed Matter Physics and Materials Science Department, Brookhaven National Labs, Upton, New York, 11973 USA*

[4]*AIST Tsukuba Central 2, 1-1-1 Umezono, Tsukuba, Ibaraki 3058568, Japan*

**A Fermi arc**[1,2] is a disconnected segment of a Fermi surface observed in the pseudogap phase[3,4] of cuprate superconductors. This simple description belies the fundamental inconsistency in the physics of Fermi arcs, specifically that such segments violate the topological integrity of the band[5]. Efforts to resolve this contradiction of experiment and theory have focused on connecting the ends of the Fermi arc back on itself to form a pocket, with limited and controversial success[6,7,8,9]. Here we show the Fermi arc, while composed of real spectral weight, lacks the quasiparticles to be a true Fermi surface[5]. To reach this conclusion we developed a new photoemission-based technique that directly probes the interplay of pair-forming and pair-breaking processes with unprecedented precision. We find the spectral weight composing the Fermi arc is shifted from the gap edge to the Fermi energy by pair-breaking processes[10]. While real, this weight does not form a true Fermi surface, because the quasiparticles, though significantly broadened, remain at the gap edge. This non-quasiparticle weight may account for much of the unexplained behavior of the pseudogap phase of the cuprates.

In a solid the behavior of the electrons is most fully described in terms of the electron Green's function, the poles of which map the energy vs. momentum dependence of the electronic quasiparticles (the dressed electronic states)[5]. The locus of poles at the Fermi energy, $E_F$, defines the Fermi surface of the material, from which almost all of the electronic properties of a material emanate. For a single continuous band, the Fermi surface should form a continuous loop. Consequently, the broken segments of Fermi surface known as Fermi arcs apparently require a major rethinking of some of the basic tenets of condensed matter physics.

One approach to resolving this problem is completely discarding the notion of electron quasiparticles, and with it almost all of the understanding of solids built up from generations of condensed matter physicists. Much support for this line of reasoning came from angle resolved photoemission spectroscopy (ARPES), which is unique in its ability to directly probe the electronic excitations as a function of energy and momentum, i.e. the quasiparticles. Earlier ARPES studies found that the ARPES peaks were either anomalously broad or vanishingly weak [11,12,13] especially in the underdoped "pseudogap" regime of the cuprates – aspects that were widely taken as evidence for the lack of electron-quasiparticles. The recent introduction of laser and low energy ARPES[14] made tremendous advances in the peak sharpness and spectral weight at $E_F$, handicapping this line of argument. We for example now see sharp nodal quasiparticle-like peaks for all doping levels of $Bi_2Sr_2CaCu_2O_{8+\delta}$ (Bi2212) studied (down to moderately underdoped x=.10 samples – see Fig. 3b black), in contrast to recent studies which were unable to observe peaks below x=0.20[13]. Therefore other ideas are needed to understand the origin of the unusual non-Fermi liquid behavior in the cuprates. Here we show that regardless of whether the peaks are sharp enough to be considered quasiparticles, these states do not reach the Fermi energy and so cannot dominate the transport and thermodynamics the way that regular quasiparticle states normally would. This finding drops the relevance of the question of whether these states are sharp enough to be quasiparticles. Instead, we argue that a more relevant issue for understanding the nature of the non-Fermi liquid behavior is whether these excitations reach the Fermi energy at all.

In addition to its ability to measure quasiparticle peaks, the momentum-resolution of ARPES makes it a preferred tool to study gaps and Fermi surfaces, and therefore Fermi arcs. Despite its great power and directness, ARPES has lacked appropriate quantitative analysis techniques, which can also significantly affect the qualitative picture that emerges. For example, gap values have been predominantly measured using the approximate "midpoint of leading edge"[15] or "peak separation of symmetrized spectra"[1,2] techniques, each of which is known to fail in many limits. Phenomenological models[16] have also been used to fit gapped energy distribution curves (EDCs)[17], but since the EDC lineshape is not yet understood[18], these fittings are not obviously better than the approximate measures.

To avoid the difficulties inherent in the traditional techniques of analyzing ARPES spectra, we have developed and employed a new method, the tomographic density of states (TDoS), the creation of which is illustrated in Fig 1, with more details in the supplementary materials. Briefly, the TDoS is a one-dimensional momentum-sum of the coherent electronic spectral weight (black curve in D), which is then normalized to a similar but ungapped reference momentum-sum along the node (black curve in B), with some similarities to the proposal by Vehkter & Varma[19]. The resulting TDoS (panel F) represents the density of states and is in many ways equivalent to a typical Giaever tunneling curve of an s-wave superconductor[20], except that it is localized to a single slice through the band structure (hence the name tomographic, meaning sliced or sectioned). This unique advantage is indispensible when electron interactions are strongly momentum dependent (e.g. in a d-wave superconductor).

To analyze the TDoS, we use the formula first proposed by Dynes to explain tunneling spectra from strongly coupled s-wave superconductors:

$$I_{TDoS}(\omega) = \rho_{Dynes}(\omega) = \text{Re} \frac{\omega - i\Gamma_{TDoS}}{\sqrt{(\omega - i\Gamma_{TDoS})^2 - \Delta^2}}$$

(1)

where $\omega$ is the energy relative to $E_F$, $\Gamma_{TDoS}$ is the pair-breaking scattering rate and $\Delta$ is the superconducting gap[21,22]. This formula is essentially a simple BCS density of states with gap $\Delta$, broadened by the parameter $\Gamma_{TDoS}$, which is interpreted as the rate of pair-breaking. This formula has been extensively tested on conventional superconductors[23], and while initially phenomenological, it has been derived from the Eliashberg theory[22]. Since each TDoS is specific to a single location on the Fermi surface the gap magnitude is single valued, allowing us to use Dynes' original form even in d-wave superconductors. An example fit of a TDoS to (1) is shown in panel F, with the extracted superconducting gap value $\Delta$=5.9±0.1 meV and scattering rate $\Gamma_{TDoS}$=2.6±0.1 meV. The $\Gamma_{TDoS}$ values extracted here are consistent with what has been measured by tunneling[24] and optics[25], though previous ARPES-based methods to extract electronic scattering rates relied on the EDC or MDC widths that give values roughly an order of magnitude larger[26] – see for example the EDCs of panel A that have widths of order 20 meV. This difference arises because $\Gamma_{TDoS}$ is only sensitive to pair-breaking interactions while the EDCs and MDCs are sensitive to all electronic scattering processes (see Fig SM3).

Taking advantage of the momentum selectivity of the TDoS, Figure 2 details the evolution of the TDoS along the Fermi surface in the near-nodal regime from an optimally doped ($T_c$=91K) Bi2212 sample. By fitting to the Dynes formula (eqn 1), we find the gap is linear and symmetric about the node as expected for a d-wave gap (Fig. 2E). In this near-nodal regime, $\Gamma_{TDoS}$ is essentially isotropic and of the scale of 2-3 meV, which is counter to earlier ARPES observations in which $\Gamma$ was large (of the order of 20-30 meV) and grew rapidly away from the node[26].

In Fig. 2F, we compare the TDoS gap measurements with the EDC leading edge[15] technique (data in panel C) and symmetrized EDC[1,2] technique (panel D), all of which came from the identical ARPES data sets. We show $\Delta_{TDoS}$ is roughly the average of the "standard" techniques away from the node. However the other methods both fail in proximity to the node. Specifically, both the EDC leading edge method and the symmetrized EDC methods have a short "arc" near the node in which the extracted gap value is either zero or negative (which is often artificially set to zero). Directly observing the sharp cusp of the d-wave gap so near the node confirms the findings of other probes such as thermal conductivity[27]. By comparing Figs. 2e and 2f we see that the threshold for the artificial zero gap in the conventional techniques is when $\Gamma_{TDoS} \approx \Delta$. This zero gap regime is what previously would have been considered a Fermi arc, i.e. the zero gap regime would have indicated the presence of a real portion of Fermi surface[2].

To study the Fermi arc in greater detail we move to an underdoped sample ($T_C$=65K) with finely gridded momentum maps in the superconducting state (T=10K) and the normal (i.e. pseudogap and/or prepairing) state (T=75K) (Fig. 3). The formation of the Fermi arc is evident in the Fermi surface maps (Fig. 3A). The symmetrized EDCs show a small (2°) Fermi arc at 10K but a much larger one at 75K, with a gap from this method (determined by the depression of weight at zero energy) only definitively present at 10° and beyond (Fig. 3B). However the TDoS paint a different picture, showing a smooth evolution of the gap rather than a discrete change at any one angle (Fig. 3C). Our observation of a finite near-nodal gap above $T_C$ is only possible if there are pre-formed superconducting pairs in the pseudogap state[28,29].

Assuming a simple d-wave $\Delta$ and an isotropic $\Gamma_{TDoS}$, we can fit the entire momentum dependence of the TDoS at once (Fig. 3D). Both the superconducting and pseudogap states are well fit with this form. While $\Delta_{Max}$ has shrunk by 30% from the superconducting state to the normal state, $\Gamma_{TDoS}$ has more than doubled, completely filling in the gap for states close to the node. These filled-in states are the source of previous observations of a Fermi arc and should be considered a manifestation of superconductive pre-pairing in the pseudogap state. The extremely short arc in the superconducting state is most likely due to in-plane impurity scattering (e.g. Cu vacancies).[30,31] However, the growth of the arc with temperature cannot be explained by static disorder. Instead, a dynamical process must lead to the increased pair-breaking rate that causes the arc to grow as was first proposed by Norman et al.[10]

A more complete temperature dependence of the Fermi arc is shown in Fig. 4A, for a similarly underdoped sample ($T_C$=67K). The growth of the Fermi arc with temperature[2] can be accurately simulated with an electron Green's function which utilizes a tight-binding band structure obtained from experiment, a simple d-wave gap ($\Delta_{Max}$=38meV), and a scattering rate $\Gamma_{TDoS}$ which is constant across the Fermi surface (Fig. 2E) but which varies strongly with temperature. The values of the electron scattering rate are directly obtained from the Dynes fits up to 110K. At higher temperatures the fits become less reliable, so we have extrapolated those values. While the gap is seen to close slightly with increasing temperature (Fig. 3D), it was held constant in Fig. 4 for simplicity. The observed growth of the Fermi arc with increased temperature can be understood with the schematic of Fig. 4B. As the dominant effect of raising temperature is to increase $\Gamma_{TDoS}$, the threshold of $\Gamma_{TDoS}=\Delta$ moves away from the node increasing the arc length.

The notion of discontinuous Fermi arcs put forward from the previous ARPES experiments is unphysical within the context of standard condensed matter theory, because Fermi surfaces must be topologically connected. Therefore, significant effort has been expended to observe if and how these arcs close to form a Fermi pocket. For example, recent ARPES experiments[6,7] have presented evidence for arc closures, though many of these have been met with skepticism for reasons including potential contamination from shadow bands, superstructure bands[9], or extended extrapolations. With the understanding presented here, there is no physical reason why the arc would need to be closed because it is not a real locus of quasiparticle states. In this light, the difficulty in observing the "back" side of the arc is completely natural – that is, the arc is actually arc-like, as opposed to only being the front side of a small hole pocket.

High magnetic field quantum oscillation studies in the cuprates have found evidence for a small Fermi surface pocket[8], with these pockets potentially related to the pockets discussed in some of the ARPES experiments. However, Hall effect measurements indicate that the pockets hold electrons instead of holes[8], making it clear that the pockets observed in the quantum oscillation experiments are inconsistent with the ARPES Fermi arc with back-side closure. Additionally, new evidence from NMR indicates that the high-magnetic fields used in the quantum oscillation experiments cause a reconstruction of the Fermi surface that can create the small pockets observed in those experiments[32].

Within the conventional analysis method of symmetrized EDCs, the Fermi arcs are considered loci of quasiparticle states at the Fermi level that grow with increasing temperature. Instead, the above simulations and data argue that the arcs are a result of the interplay between $\Delta(k,T)$ and $\Gamma_{TDoS}(k,T)$, which we find is dominated by $\Delta(k)$ and $\Gamma_{TDoS}(T)$. Importantly, this result indicates that the arcs are not composed of true quasiparticles because, even though the quasiparticles exist, they are not at at $E_F$ – rather the pole locations are the gap values, $\Delta(k)$ (Fig 4C). Rather than a true Fermi surface of quasiparticle poles, the arcs are regions where real spectral weight has been scattered inside the d-wave gap, with this incoherent non-quasiparticle weight varying with Fermi surface angle and temperature (Fig. 4C). Aside from the single quasiparticle state at the node, this non-quasiparticle weight comprises the only states available to contribute to the low energy transport and thermodynamic properties. Therefore, these states are strong candidates to explain the unusual transport, heat capacity, and other thermodynamic properties in the pseudogap state of the cuprates[33].

**Acknowledgements.** We thank G. Arnold, A. Balatsky, I. Mazin, S. Golubov and M. Hermele for valuable conversations and D. H. Lu and R. G. Moore for help at SSRL. SSRL is operated by the DOE, Office of Basic Energy Sciences. ARPES experiments at the Hiroshima Synchrotron Radiation Center were performed under proposal 09-A-48. Funding for this research was provided by DOE Grant No. DE-FG02-03ER46066 (Colorado) and DE-AC02-98CH10886 (Brookhaven) with partial support from the NSF EUV ERC and from Kakenhi (10015981 and 19340105).

**Contributions**
T. R., N. P., Z. S., Y. C., Q. W., M. A. and H. I. performed the ARPES measurements. J. W., Z. X., G. G., Y. Y. , H. E. and Y. A. grew and prepared the samples. T.R. analysed the ARPES data. T.R and D.D. developed the TDoS technique and wrote the paper with suggestions and comments by N.P., Z. S. and K. M. ARPES simulations were done by T. R. D.D. is responsible for project direction, planning and infrastructure.

.
Correspondence and requests for materials should be addressed to D. D. (Dessau@Colorado.edu)

**Fig 1: Creation and Fitting of the Tomographic density of states (TDoS). a)** Nodal spectrum of optimally doped ($T_C$ =92K) Bi2212 at T=50K.  **b)** (colored) 7 out of a total of 170 individual energy distribution curves (EDCs) taken along the horizontal colored lines of panel a. The green EDC corresponds to k=$k_F$. The sum of all 170 individual EDCs gives the spectral weight curve (black). **c)** Slightly (6°) off nodal spectra taken along the red cut in the inset. **d)** EDCs and the corresponding spectral weight for the data of panel C showing the superconducting gap at $E_F$ as well as additional spectral weight above $E_F$ due to the Bogoliubov quasiparticles. **e)** Off-nodal spectra normalized to nodal spectral weight (color scale data) and the spectral weight curves from panels B and D (orange and red curves, respectively).  The red curve is normalized to the orange curve to create the tomographic density of states or TDoS curve (Yellow). **f)** Fit of the TDoS (red circles) to the Dynes tunneling formula (green line).

**Fig 2: Angular dependence of the TDOS and comparison to conventional ARPES analysis techniques from an optimal $T_C$=91K Bi2212 sample at T=50K. a)** Angle dependence of ARPES spectra. **b)** TDoS curves vs. angle. **c)** EDC's at $k_F$ vs. angle. **d)** Symmetrized EDC's at $k_F$ vs. angle, all from the same data set. **e)** $\Delta_{TDoS}$ and $\Gamma_{TDoS}$ for angles very near the node. Error bars are ±σ returned from fits to the data of panel B. **f)** $\Delta_{TDoS}$ near the node compared to Δ's measured by the other two methods.  These other methods show a short but artificial "arc" of zero gapped states near the node, while $\Delta_{TDoS}$ extrapolates to 0 meV only at the node.

**Fig 3: Comparing the TDoS and Symmetrized EDCs when determining a Fermi Arc, for an underdoped ($T_C$=65K ) sample. a)** Spectral weight at $E_F$ both in the superconducting (T=10K, top) and pseudogap (T=75K, bottom) states. **b)** Symmetrized EDCs from which one would determine a finite range of gapless states at $E_F$ that grows with increasing temperature. **c)** Measured TDoS spectra as a function of angle away from the node. **d)** Two parameter fits ($\Gamma_{TDoS}$ and $\Delta_{Max}$) for each entire set of spectra, with parameters listed on the panels. $\Delta_{max}$ is the max of the d-wave gap, occurring at the antinode, with the k-dependence of the gap forced to maintain the simple d-wave form. $\Gamma_{TDoS}$ is held to a constant value throughout k-space, consistent with the results of fig 2E.

**Fig 4: Simulating the observed temperature dependence of the Fermi arc from a $T_c$=67K underdoped $Bi_2Sr_2CaCu_2O_8$ sample. a)** The unusual temperature-dependent arc length is seen in both the data and simulations.  The main inputs to the simulations are an experimentally determined tight binding band structure, a simple d wave gap (both below and above $T_c$) with independently measured d-wave gap maximum 40 meV, and a temperature dependent scattering rate $\Gamma_{TDoS}$ (shown) with units of meV.  The majority of these $\Gamma_{TDoS}$ values (<110K) are also independently experimentally determined. **b)** Simple picture for the temperature dependence of the Fermi arc length.  The arc is approximately the region where $\Gamma_{TDoS} > \Delta$, which grows rapidly with temperature as $\Gamma_{TDoS}$ grows. **c)** Cartoon showing change in spectral weight (red) along the Fermi surface due to increasing $\Gamma_{TDOS}$ between the superconducting and normal states as determined by the TDoS (green)

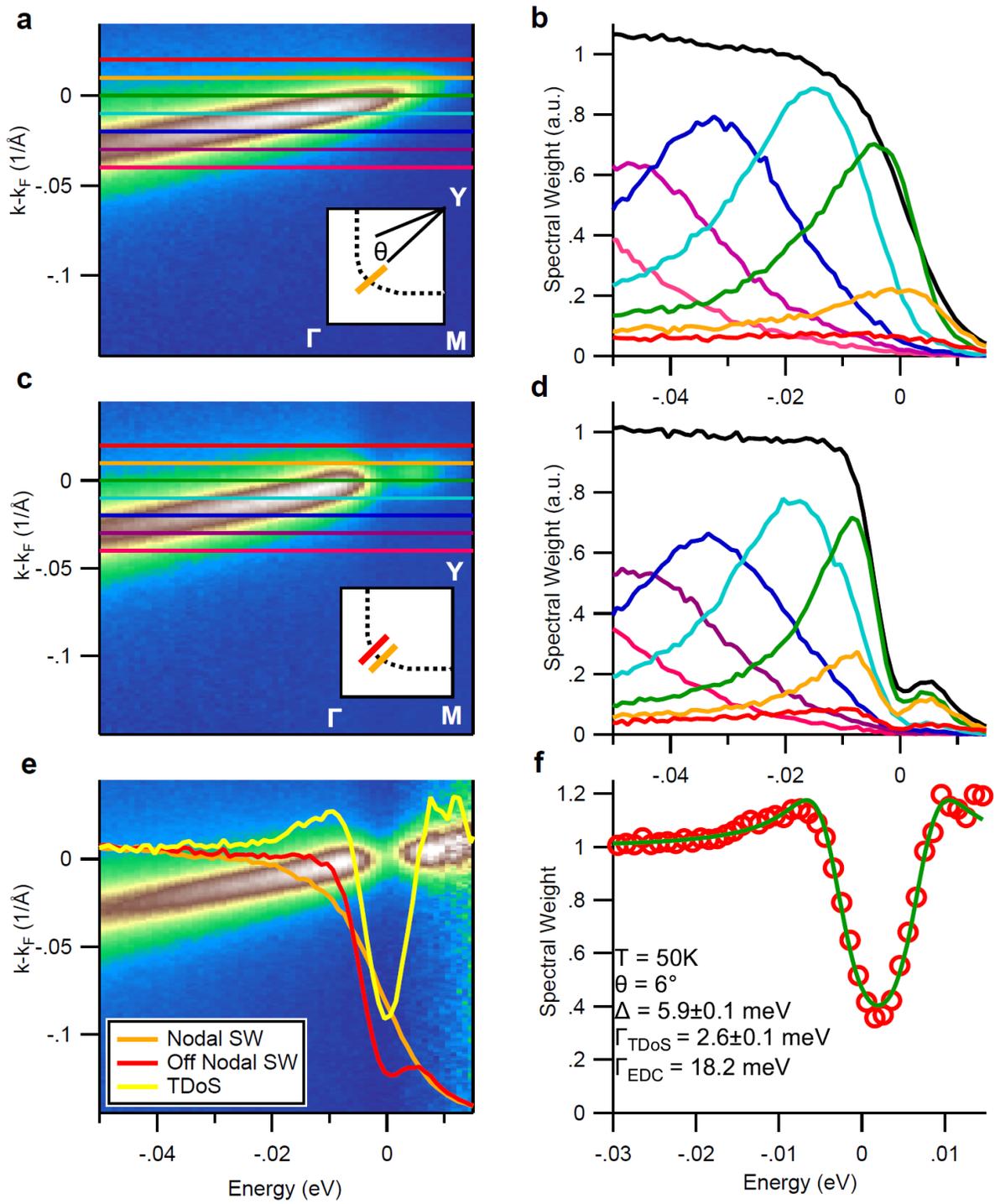

Figure 1

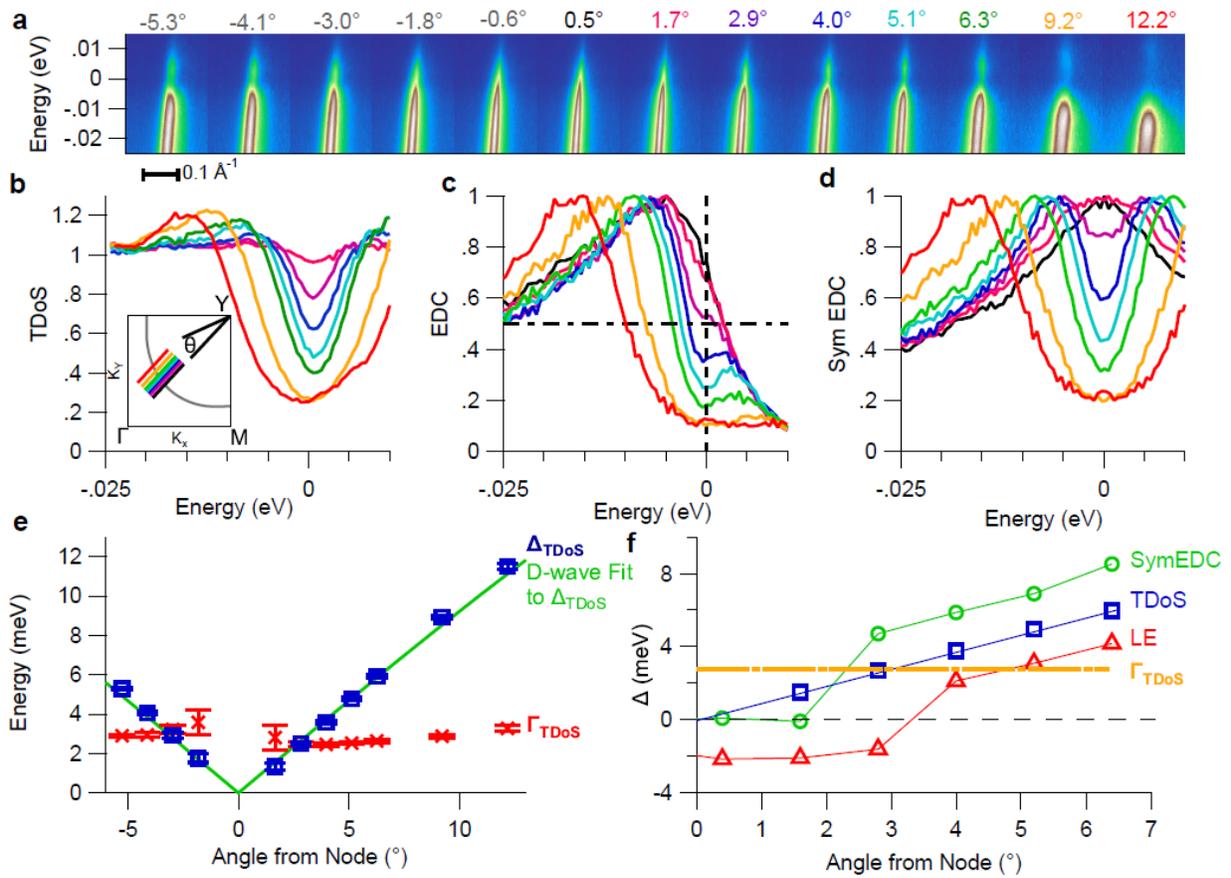

**Figure 2**

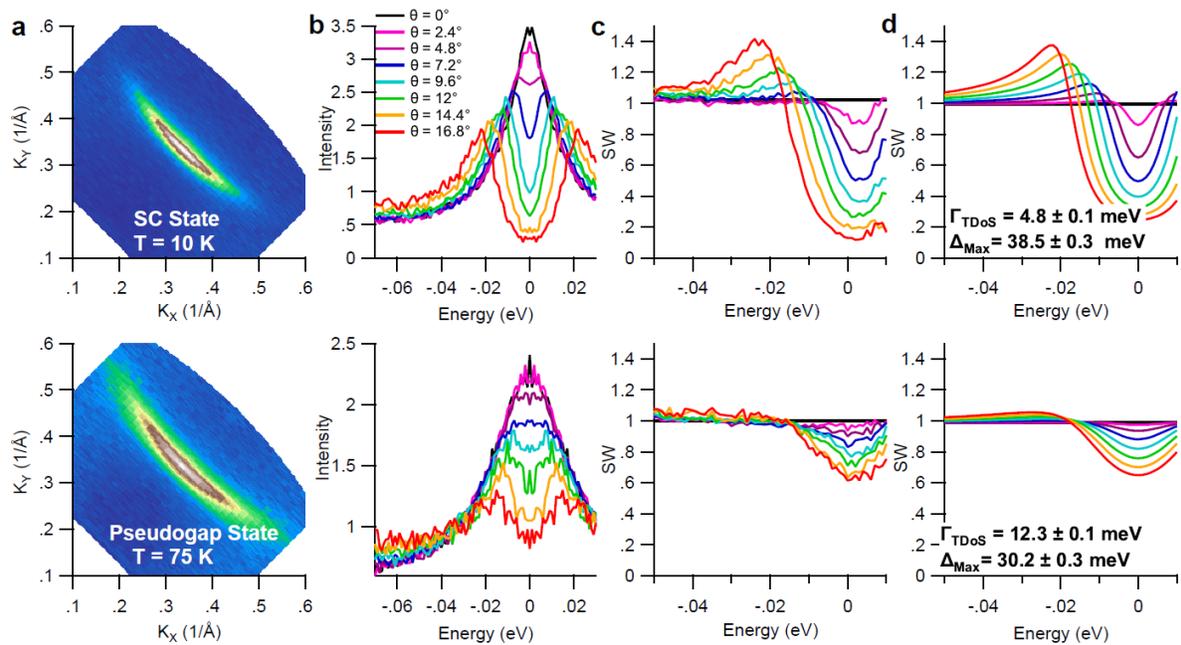

**Figure 3**

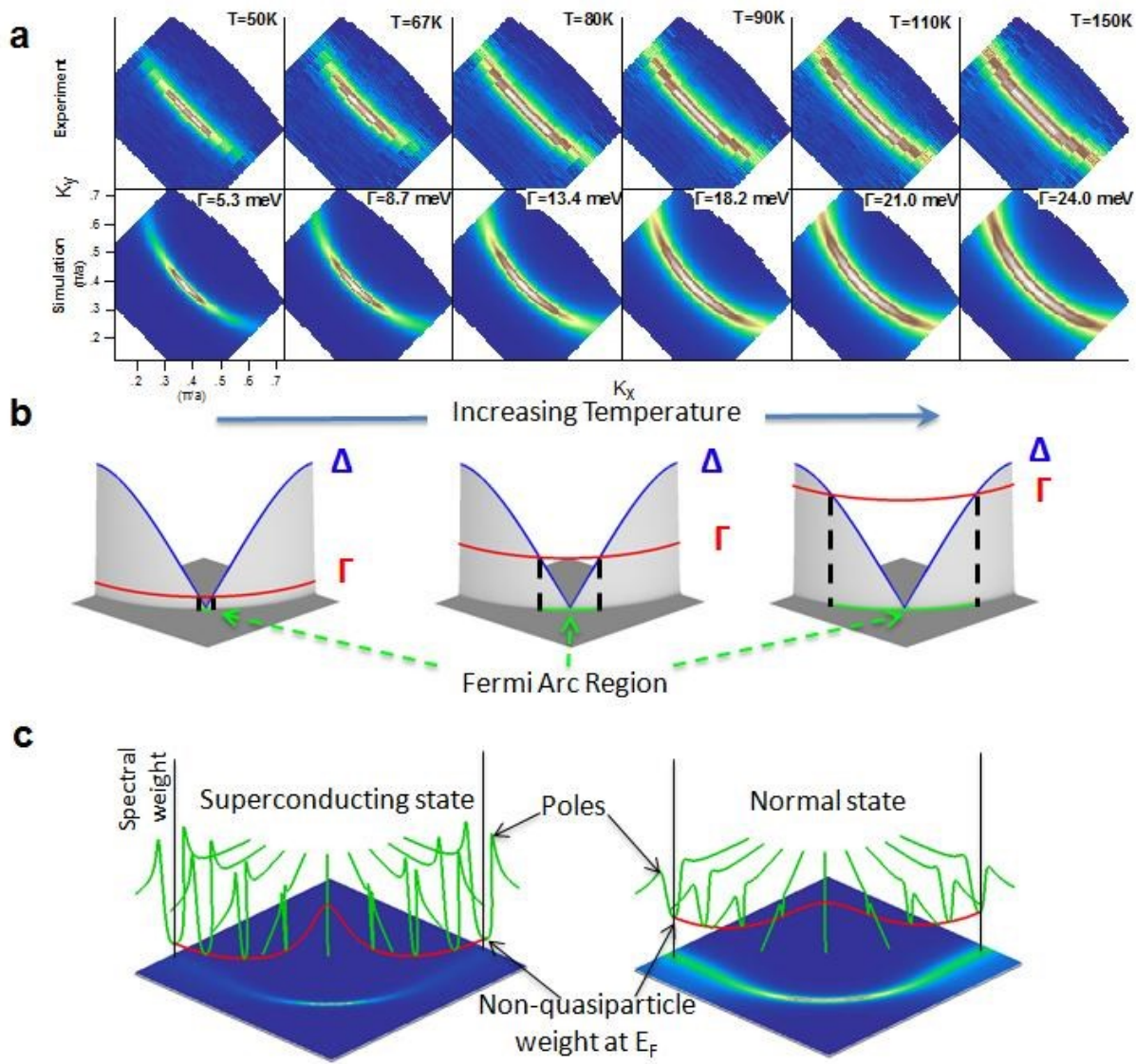

**Figure 4**

**Supplemental Methods**

**Samples and experimental equipment.** The samples studied in this letter were high quality $Bi_2Sr_2CaCu_2O_{8+\delta}$ single crystals cleaved in ultra-high vacuum. All data were taken with a Scienta R4000 analyzer using 7 eV photons. The Scienta electron detection systems have a noticeable nonlinearity as a function of counting rate that can be up to a 30% increase[34,35]. To our knowledge this effect has so far been ignored in the analysis of ARPES spectra though it can play a significant role, especially near the Fermi edge where spectral weight is rapidly decreasing. We have calibrated and corrected for this effect. All other methods of experimentation are standard for the field. TDoS's do not require very high angular resolution, though high energy resolution is important because we will be studying superconducting gaps of the order of a few meV. For this reason we use low energy photons (7 eV), giving an experimental resolution of approximately 4 meV FWHM.

**Data treatment- background removal and spectral weight determination.**

We find it is very helpful to isolate the weight of the dispersive states from that of the background non-dispersive states. We have done that here by analyzing each spectrum via momentum distribution curves (MDCs). We fit each MDC to a Lorentzian with a constant offset, taking the area of the Lorentzian to be the dispersive spectral weight and the offset to be the background. Figure SI1a shows a sample MDC from a nodal cut at 50K (red), its fit (teal), the background weight (blue) and the band weight (green). The energy dependence of these various spectral weights is shown in Fig SI1b, while the effect of the background removal on the overall spectrum is shown in Fig SI1c. The BG-removed spectrum looks essentially identical to the original spectrum, except that some of the haze around the dispersive band-like states is missing.

**Including Energy Resolution in Fitting the TDoS,**

Though small in low energy ARPES, the energy resolution is finite and for the most accurate fits we need to include it. To properly include the energy resolution we can't simply convolve the Dynes formula with a Gaussian with the width of the energy resolution. That method fails to consider the effect of the rapidly decreasing weight from the Fermi Edge. Instead we must consider how the experimental resolution affects the data prior to the normalization procedure. Consequently, we use the formula:

$$Fit(\omega) = \frac{(F(\omega)\rho_{Dynes}(\omega)) \otimes R(\omega)}{F(\omega) \otimes R(\omega)} \qquad (S1)$$

Where $F(\omega)$ is the Fermi Function, $\rho_{Dynes}(\omega)$ is the Dynes formula, $R(\omega)$ is the experimental resolution. The main effect of including the resolution is the deviation from the expected "U" shape inside the gap to the "V" shape of our TDoS. Once the energy resolution is included the quality of the fits is excellent as shown in Fig. SI2, where we have reprinted the TDoS from Fig 2b (Fig. SI2).

**Comparing $\Gamma_{TDoS}$ to $\Gamma_{MDC}$**

The standard ARPES method for determining the scattering rate, momentum distribution curve (MDC) widths[36], gives static values $\Gamma_{MDC}(\omega=0, T\sim0)$ nearly an order of magnitude larger than the associated values for $\Gamma_{TDOS}$. Note that this data is from low photon energy ARPES, which as first shown by Koralek et al. returns $\Gamma_{MDC}$ significantly lower than standard ARPES experiments[37]. Furthermore, while the TDoS method returns almost identical scattering rates from many samples (Fig. SI3a), the $\Gamma_{MDC}$'s from the same exact samples vary by more than a factor of 2 (Fig. SI3b). This wide variation in MDC widths without a similar variation in the $T_C$ indicates that the MDCs (and EDCs) are subject to scattering processes that do not affect the TDoS's (or the pairing). These results show that the physical processes that are responsible for the MDC broadening $\Gamma_{MDC}$ are different from (or in addition to) those that are responsible for the TDoS broadening $\Gamma_{TDoS}$.

To understand the difference between the $\Gamma_{MDC}$ and $\Gamma_{TDoS}$, we plot the effect of increasing $\Gamma_{TDoS}$ on density of states using Dynes formula, in Fig. SI3c. The dominant effect of increasing $\Gamma$ is to shift weight from the peak at $\Delta$ up to $E_F$. Since we are restoring weight to the Fermi energy, we argue $\Gamma_{TDoS}$ represents only the scattering processes that can break pairs while $\Gamma_{MDC}$ represents all scattering processes. For example, In a d-wave superconductor we expect pair-breaking scattering to be those scattering events with a $q=k'-k$ which change the phase of the order parameter, whereas non-pair-breaking scattering would not change the phase and are likely related to highly prevalent forward scattering events[38]. Figure SI3d schematically illustrates both a pair-breaking (blue) and non-pair-breaking scattering event (red). We imagine that out-of-plane disorder would be well screened and would predominantly cause forward scattering and not break the pairs, while in-plane disorder (e.g. Cu vacancies) would give the large q scattering necessary for pair breaking[39,40].

**Simulating Nodal and Off-Nodal Spectra**

Using the electron Green's function, it is simple, in principle, to recreate our observed ARPES spectra. Including the Fermi Edge, energy dependent self-energy and experimental resolution, the ARPES spectra can be written:

$$I_{ARPES}(\omega,k) = \left( F(\omega) \times -\frac{1}{\pi} \mathrm{Im} G(\omega,k) \right) \otimes R(\omega,k) \tag{S2}$$

Where $F(\omega)$ is the Fermi function $R(\omega,k)$ is the experimental resolution and the Green's function is written:

$$G(\omega,k) = \frac{\omega + \Sigma - \varepsilon_k}{(\omega + \Sigma)^2 - (\varepsilon_k)^2 - \Delta_k^2} \tag{S3}$$

where $\Sigma$ is the complex self energy and $\varepsilon_k$ is the bare band and $\Delta$ is the gap magnitude. For these simulations, we used a reasonable interpretation that Im($\Sigma$) linear and symmetric about $E_F$ with a step at 70 meV and the minimum value as determined from the MDC width (20 meV). We took the Kramers-Kronig transform of Im($\Sigma$) to extract the Re($\Sigma$). We assumed a linear bare band with $v_{BB}$ = 2.9 eV/Å and $k_F$ = .42 1/Å. Finally, we set T to 50K and the energy resolution at 4 meV. For the nodal ($\Delta$=0) case this simulation does an excellent job of recreating the observed spectrum as shown in Fig SI4. To simulate the slightly off-nodal spectra we should only have to add the presence of a small but finite $\Delta$. However, the addition of a finite $\Delta$ has no observable effect on the on the simulation, even though the gap is clearly resolvable in the data. We find that the minimum Im($\Sigma$) is just too large to achieve a significant depreciation in weight at the Fermi energy for the off-nodal spectrum. We can achieve the expected reduction in spectral weight by drastically reducing the minimum Im($\Sigma$) by an order of magnitude. However the band's width would be dramatically too narrow if we only used $\Gamma_{TDoS}$. Instead, we can compensate by broadening the spectrum as if the MDC width was dominated by extrinsic final state effects by convolving a Lorentzian in momentum with the spectrum. The resulting simulation matches the observed spectra quite well for both the gapped and ungapped cases. This result further supports our conclusion that $\Gamma_{TDoS}$ more accurately describes the superconducting state scattering rates than $\Gamma_{MDC}$.

**Creating and Simulating the Fermi Surface Maps**

To create a Fermi surface map that is representative of the coherent spectral weight at $E_F$ and thus the Fermi arc, we found it necessary to remove the incoherent background and correct for the photon energy dependent matrix elements. We removed the background as described previously for each individual angle. For 7 eV photons, the suppression of spectral weight due to matrix elements is dramatic beyond 10° from the node. If left uncorrected this effect would artificially shrink the observed Fermi arcs. To compensate, we normalize the coherent spectral weight for each angle over the range -30 to -70 meV, well below the region of the gap.

To simulate the Fermi surface maps in Fig. 4, rather than the spectral function, we started with electron Green's function of the form

$$G(\omega, k_x, k_y) = \frac{\omega + i\Gamma_{TDoS} - \varepsilon_k}{(\omega + i\Gamma_{TDoS})^2 - (\varepsilon_k)^2 - \Delta_k^2} \quad (S4)$$

Where $\Gamma_{TDoS}$ is the pair-breaking scattering rate, $\Delta_k$ is a d-wave gap of the form

$$\Delta_k = \frac{\Delta_0}{2}(\cos k_x - \cos k_y) \quad (S5)$$

And $\varepsilon_k$ is the band structure, a tight binding model of the form

$$\begin{aligned}\varepsilon_k &= u - 2t_0(\cos k_x + \cos k_y) - 4t_1(\cos k_x \cos k_y) - 2t_2(\cos 2k_x + \cos 2k_y) \\ &\quad - 4t_3(\cos 2k_x \cos k_y + \cos k_x \cos 2k_y) - 4t_4(\cos 2k_x \cos 2k_y)\end{aligned} \quad (S6)$$

Extracting just the imaginary part, we recover the spectral function $A(\omega, k_x, k_y)$ which ARPES actually measures. To account for the large MDC widths, we broaden the spectra as previously discussed for the 2-D simulations, but along both $k_x$ and $k_y$ for the 3-D case.

**Supplemental Discussion**

**Mathematical Comparison of TDoS to Normal ARPES and Tunneling**
As the TDoS is a merger of ARPES data with tunneling analysis, we find illustrative to directly compare how ARPES, tunneling and the TDoS measure the spectral function $A(k,\omega)$. Rather than the usual $k_x$ and $k_y$, we define two orthogonal momenta, $k_{ll}$ and $k_\perp$, where $k_{ll}$ is the momentum parallel to the Fermi surface while $k_\perp$ is normal to the Fermi surface, though still in the *xy* plane ($k_z$ would be the out-of plane direction). $A(k,\omega)$ is the spectral function, $M^2(k,\omega)$ is the ARPES or tunneling matrix element, and $f(\omega)$ is the Fermi function.

The intensity measured by ARPES (ignoring experimental resolution) is:

$$I(k,\omega)_{ARPES} = A(k,\omega)M^2(k,\omega)f(\omega) \tag{S7}$$

While $I(k,\omega)_{ARPES}$ contains the most detail, the pairing processes can be swamped by non-thermodynamic scattering processes like forward scattering. Tunneling experiments integrate over all momenta such that:

$$I(\omega)_{tunneling} = \int_{k_\perp}\int_{k_{ll}} A(k,\omega)M^2(k,\omega)d^2k \tag{S8}$$

While tunneling is sensitive to the pairing processes, $I(\omega)_{tunneling}$ is complicated by strong variation along $k_{ll}$ such as to the d-wave superconducting gap, the van-Hove singularity near the antinode, etc. However to create the TDoS we integrate over the momentum, $k_\perp$:

$$I(k_{ll},\omega)_{TDoS} = \int_{k_\perp} A(k,\omega)M^2(k,\omega)f(\omega)dk_\perp \tag{S9}$$

By normalizing the spectra properly we obtain the simple functional as shown below,

$$\cong \int_{k_\perp} A(k,\omega)dk_\perp \tag{S10}$$

This form allows us to probe the thermodynamic pair-forming and pair-breaking processes with the use of well developed tunneling analysis techniques for a simple (i.e. s-wave) gap[41], even though the gap varies as a function of $k_{ll}$ due to the d-wave nature of the order parameter.

**Fig SI1: Removing the Background, a)** Sample MDC (red) 100 meV from $E_F$, its fit (teal), background weight (blue), and band weight (Green). The MDC at $E_F$ (dashed) is shown for comparison. **b)** Energy dependence of spectral weights for a nodal spectrum at 50K. **c)** Comparison of the original nodal spectrum and with the background removed.

**Fig SI2: Experimental TDoS spectra and Fits.** TDoS and fits as a function of angle near the node used in figure 2.

**Fig SI3: Broadening mechanisms for MDC curves not present in TDoS spectra. a)** TDoS spectra at θ=6° from 3 different samples, including extracted parameters Δ and $\Gamma_{TDoS}$. **b)** Nodal MDCs at $E_F$ from the same three samples as studied in panel a, showing the significant deviation in MDC widths from sample to sample. The extracted broadening parameters $\Gamma_{MDC}$ are shown in a. **c)** Illustration of how increasing $\Gamma_{TDoS}$ shifts weight from the peak to $E_F$ **d)** Diagram of the Fermi surface detailing the difference between non-pair-breaking forward scattering (red) and pair-breaking unitary scattering (blue).

**Figure SI4: Simulations of Nodal and Off-Nodal Spectra from the Electron Green's Function.** The first column shows experimental data in the superconducting state for the node (top and slightly off-node (bottom). The second column shows simulations using Eq. S2 and S3 with a Γ=20 meV (obtained from MDC analysis). While nodal case closely matches the experimental data, the off-nodal simulation cannot recreate the gap. Instead we must lower Γ by an order of magnitude to achieve the observed depletion of weight at $E_F$ (third column). To recover the experimental MDC widths, we then broaden the spectra by convolving a Lorentzian with a width, $\Gamma_k$, in momentum (fourth column).

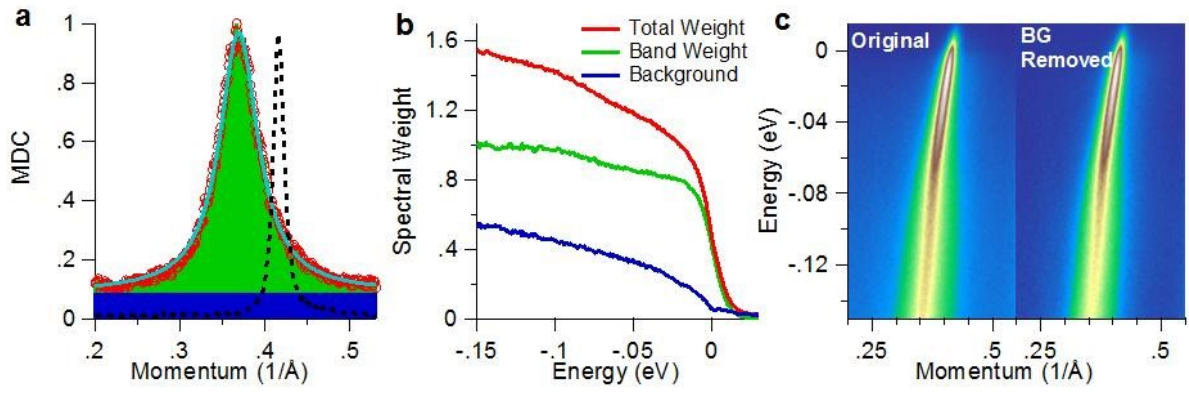

**Figure SI1**

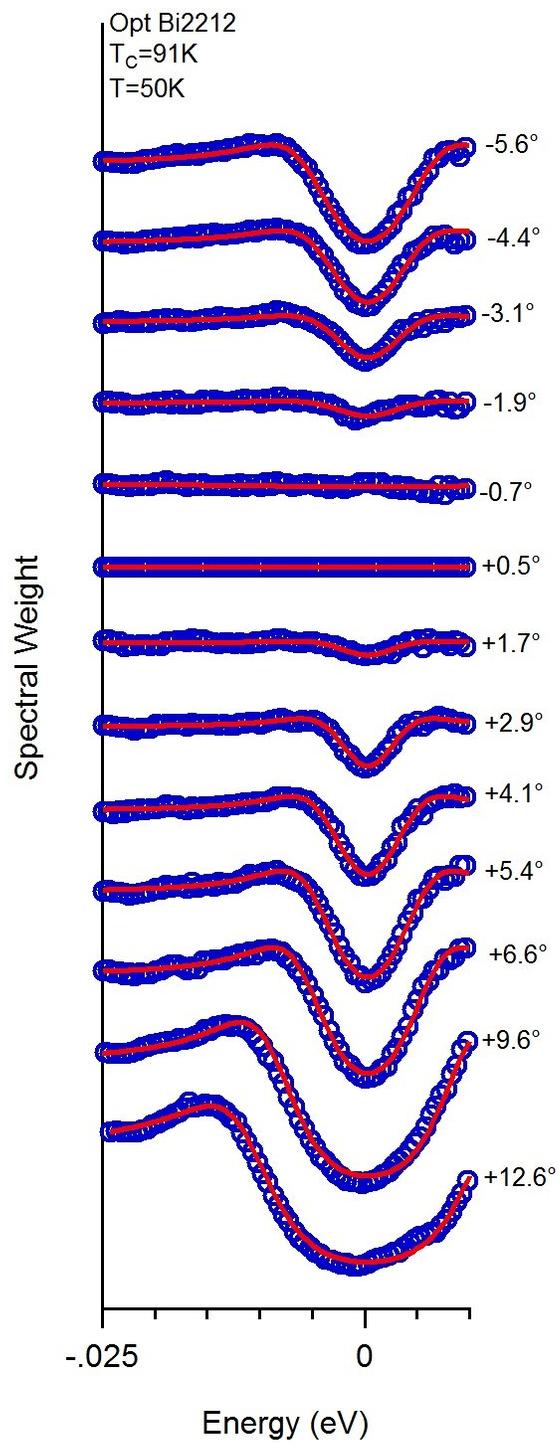

**Figure SI2**

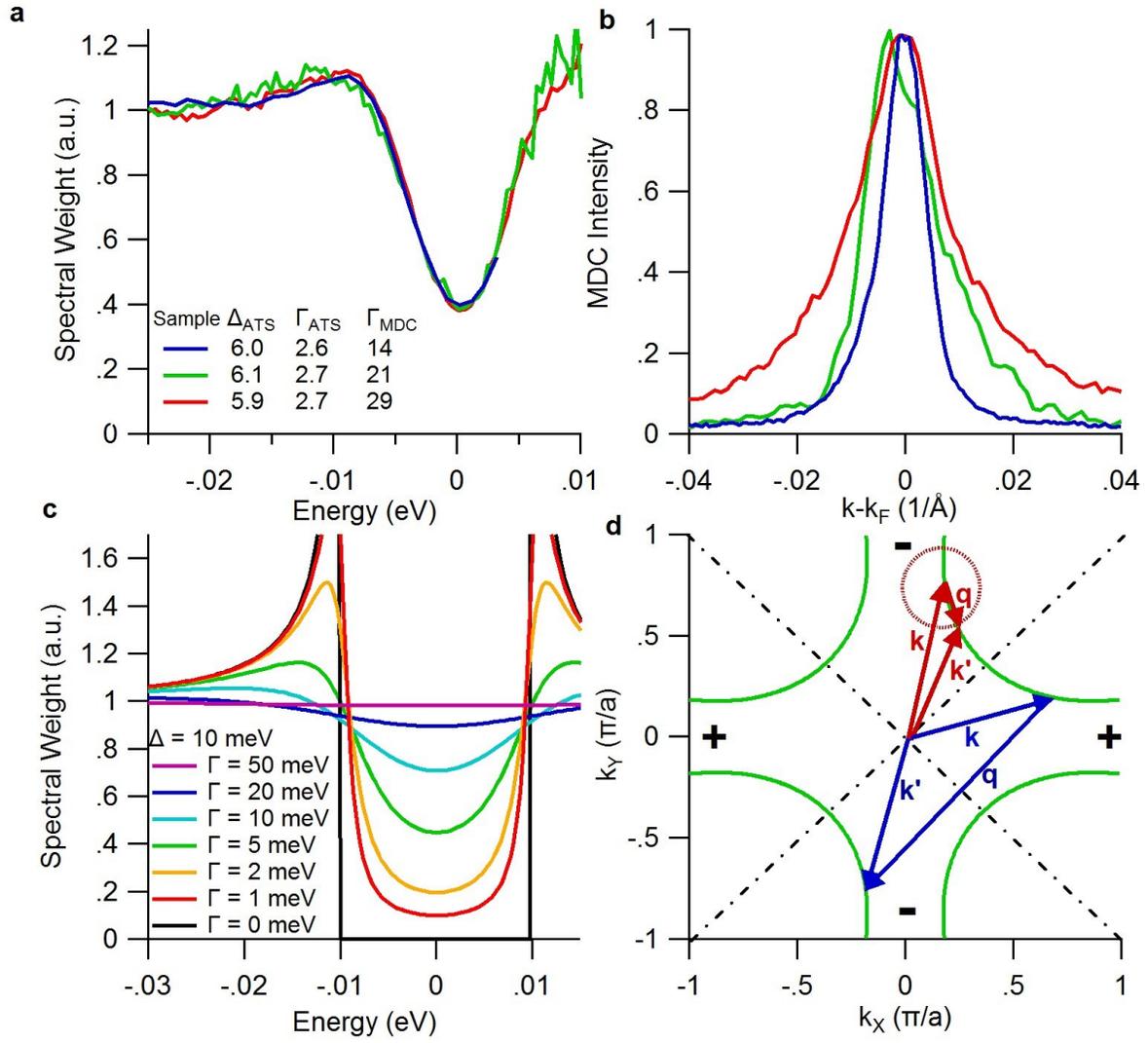

**Figure SI3**

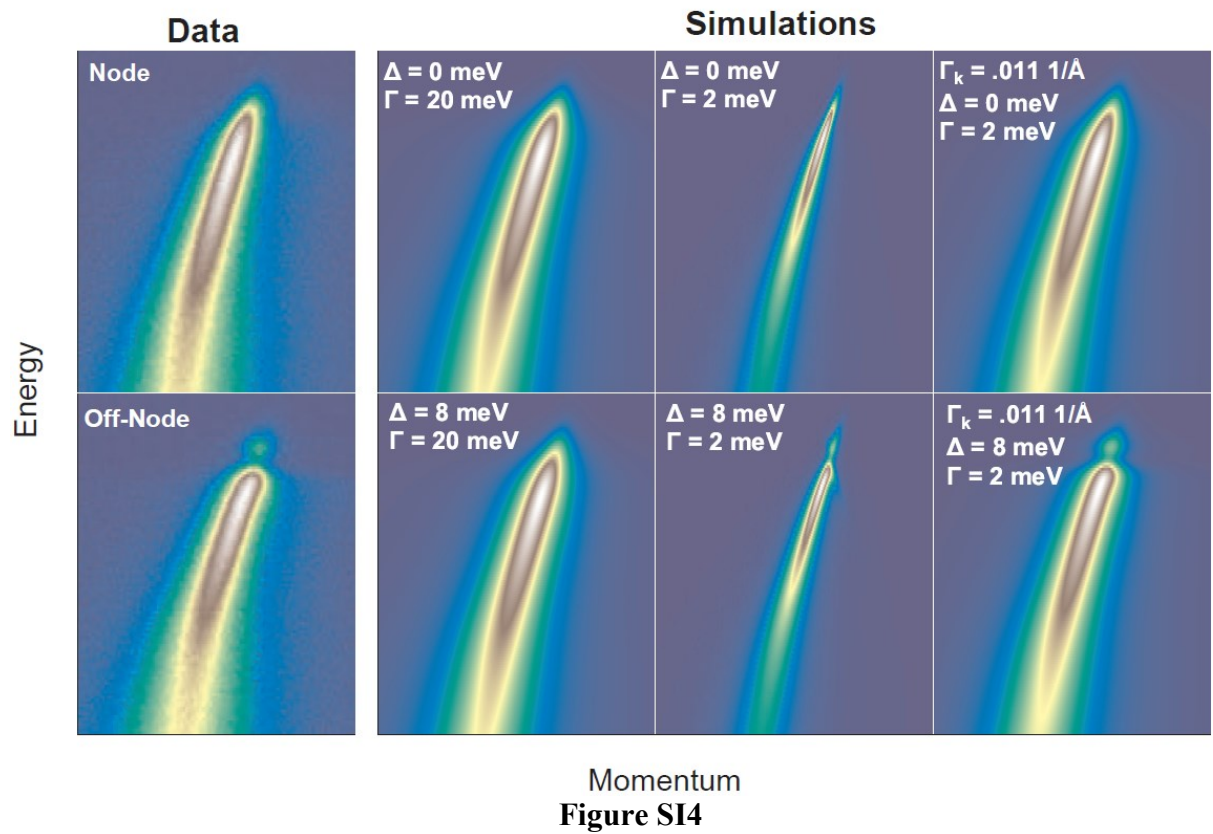
**Figure SI4**